\documentclass[useAMS,usenatbib]{mn2e}
\usepackage{bm,graphics,epsfig}
\def \tprime {t^{\prime}}
\def \tpp {t^{\prime\prime}}
\def \etad {\eta_{\rm t}}
\def \kpeak {k^{\rm peak}}
\def \axx {\alpha_{xx}}
\def \axy {\alpha_{xy}}
\def \ayx {\alpha_{yx}}
\def \ayy {\alpha_{yy}}
\def \delz {\partial_{z}}
\def \del2z {\partial^{2}_{z}}
\def \Bx { \bar{B}_x }
\def \By {\bar{B}_{y}}

\def \Bhx {\hat{B}_{x}}
\def \Bhy {\hat{B}_{y}}
\def \Bhxcc {\hat{B}^{\ast}_{x}}

\def \bhxx {{\hat B}_x {\hat B}^{\ast}_x }

\def \bhyx {{\hat B}_x^{\ast} {\hat B}_y }
\def \Bhxx {\langle {\hat B}_x {\hat B}^{\ast}_x \rangle}
\def \Bhyy {\langle  {\hat B}_y {\hat B}^{\ast}_y \rangle}
\def \Bhxy {\langle  {\hat B}_x {\hat B}^{\ast}_y \rangle}
\def \Bhyx {\langle  {\hat B}_x^{\ast} {\hat B}_y \rangle}
\def \exx {{\tilde \eta}_{xx}}
\def \exy {{\tilde \eta}_{xy}}
\def \eyx {{\tilde \eta}_{yx}}
\def \eyy {{\tilde \eta}_{yy}}

\def \etaxx {\eta_{xx}}
\def \etaxy {\eta_{xy}}
\def \etayx {\eta_{yx}}
\def \etayy {\eta_{yy}}
\def \ksqr {k^{2}}
\def \kz {k}

\def \MM {{\sf\bm{M}}}
\def \AA {{\sf\bm{A}}}
\def \MMd {{\sf\bm{M}^{\rm d}}}
\def \tcor {T_{\rm corr}}
\def \cB {{\mathcal B}}

\def \Ubar {{\overline {\bm U}}}

\def \ey {{\bm e}_y}

\newcommand{\D}[2]{{\mathcal D}^{#1}_{#2}}
\newcommand{\fder}[2]{\frac{\delta #1}{\delta #2}}
\newcommand{\fdertxt}[2]{\delta (#1)/\delta #2}
\newcommand{\eq}[1]{(\ref{#1})}

\newcommand{\Eq}[1]{Equation~(\ref{#1})}
\newcommand{\Eqs}[2]{Equations~(\ref{#1}) and~(\ref{#2})}

\newcommand{\eqs}[2]{(\ref{#1}) and~(\ref{#2})}

{}

\def \Rey  {\mbox{Re}}

\def \kf  {k_{\rm f}}

\def \urms  {u_{\rm rms}}

\def \etat {\eta_{\rm t}}

\def \kf  {k_{\rm f}}

\def \urms {u_{\rm rms}}

\newcommand{\apj}{ApJ}
\newcommand{\an}{Astron. Nachr.}
\def\mnras{MNRAS}

\title[Incoherent $\alpha$--shear dynamos]
{Scaling and intermittency in incoherent $\alpha$--shear dynamo}
\author[Dhrubaditya Mitra and Axel Brandenburg]{Dhrubaditya Mitra$^{1}$\thanks{E-mail:
dhruba.mitra@gmail.com}
and Axel Brandenburg$^{1,2}$\thanks{E-mail:
brandenb@nordita.org} 
\\
$^1$NORDITA, AlbaNova University Center, Roslagstullsbacken 23,
SE-10691 Stockholm, Sweden\\
$^2$Department of Astronomy, AlbaNova University Center,
Stockholm University, SE-10691 Stockholm, Sweden}
\begin{document}

\date{\today, $ $Revision: 1.144 $ $}

\maketitle

\label{firstpage}

\begin{abstract}
We consider mean-field dynamo models with fluctuating $\alpha$ effect,
both with and without large-scale shear.  
The $\alpha$ effect is chosen to be Gaussian white noise with zero mean and
a given covariance. 
In the presence of shear, we show analytically that (in 
infinitely large domains) the mean-squared magnetic field shows exponential
growth.
The growth rate of the fastest growing mode is proportional to the shear rate.
This result agrees with earlier numerical results of \cite{YHSKRICM07}
and the recent analytical treatment by \cite{HMS11} who use
a method different from ours.
In the absence of shear, an incoherent $\alpha^2$ dynamo may 
also be possible. 
We further show by explicit calculation of the growth rate of 
third and fourth order moments of the magnetic field that the
probability density function of
the mean magnetic field generated by this dynamo is non-Gaussian. 
\end{abstract}

\begin{keywords}
Dynamo -- magnetic fields -- MHD -- turbulence
\end{keywords}
\section{Introduction}
The dynamo mechanism that generates large-scale magnetic fields in 
astrophysical objects is a topic of active research. 
Almost all astrophysical bodies, e.g., the Galaxy, or the Sun,
show presence of large-scale shear (differential rotation).
It is now well established that this shear is an essential ingredient
in the dynamo mechanism. 
This view is also supported by direct numerical simulations (DNS). 
In particular, DNS of convective flows have been able to generate
large-scale dynamos predominantly in the presence of shear
\citep{kap+kor+bra08,hug+pro09}, while non-shearing large-scale dynamos
are only possible at very high rotation rates \citep{kap+kor+bra09}.
The other vital constituent of the large-scale dynamo mechanism is
helicity of the flow which is often described by the $\alpha$ effect. 
Indeed most dynamos, including the early model by \cite{par55},
are the result of an $\alpha$ effect combined with shear. 
Of these two ingredients, shear is typically constant over the time
scale of generation of the dynamo. 
For the case of the solar dynamo, shear or differential rotation
are constrained by helioseismology. 
By contrast, measuring the $\alpha$ effect is a non-trivial exercise.
Whenever it has been obtained from DNS studies, it was found to have
large fluctuations in space and time; see e.g., \cite{BRRK08} for 
the PDF of different components of the $\alpha$ tensor and \cite{CH96} for 
the fluctuating time series of the total electromotive force,
which is however different from the $\alpha$ effect.
What effect do these fluctuations have on the properties of the
$\alpha$--shear dynamo? 
In particular, can a fluctuating $\alpha$ effect about a {\it zero} mean
drive a large-scale dynamo in conjunction with shear?
Recent DNS studies
\citep[][hereafter referred to as BRRK]{B05,YHSKRICM07,BRRK08}
suggest that the answer to this question is yes.
\cite{YHSKRICM07} have further provided compelling evidence that 
the growth rate of the large-scale magnetic energy in such dynamos scales 
linearly with the shear rate, and the wavenumber of the fastest
growing mode scales as the square root of the shear rate. 
It has been proposed that such a dynamo can also emerge
by an alternate mechanism (which has nothing to do with fluctuations
of the $\alpha$ effect) involving the interaction between shear
and mean current density. 
This mechanism is called the {\it shear--current} effect \citep{RK03}.
However recent DNS studies of BRRK as well as those of \cite{B05b}
have not found evidence in support of it,
and analytical works \citep{RS06,RK06,sri+sub09}, for small 
magnetic Reynolds number, have doubted its existence.
Furthermore, the shear--current effect does not produce the
observed scaling, namely, growth-rate scales with the shear rate squared, 
and the wavenumber of the fastest growing mode scales linearly with the 
shear rate; see Section~4.2 of BRRK.

It behooves us then to consider the interaction between fluctuating
$\alpha$ effect and shear as a possible dynamo mechanism. 
\cite{kra76} was first to propose that fluctuations of kinetic helicity
can give rise to negative turbulent diffusivities, thereby
giving rise to a dynamo that is effective only in small scales \citep{Mof78}.
However, more relevant in the present context is the work of \cite{VB97},
who have investigated (using numerical tools) an one-dimensional model
and (with analytical techniques) a simple  mean-field model
in a one-mode truncation with shear and fluctuating $\alpha$ effect.
These models they referred to as incoherent $\alpha$--$\Omega$ dynamos.
By dividing the poloidal field component by the toroidal one, they
turned the mean-field dynamo equations with multiplicative noise into
one with additive noise of Langevin type.
They drew an analogy between the behaviour of the mean magnetic field
and Brownian motion in the sense that the mean field does not grow
although the mean-square field grows 
with a growth rate that scales with shear rate to the $2/3$ power.

Over the last decade, the problem has been studied using a variety of
different approaches \citep{sok97,sil00,Fed06,pro07,KR08,SS09}.
Models with fluctuating $\alpha$ effect, which depend
on both space and time, can be divided into two categories,
one in which $\alpha$ is inhomogeneous \citep{sil00,pro07,KR08} 
and the other in which it is a constant in space.  
(There are other technical differences in the two approaches as well.)
In this paper we shall confine ourselves to the second class of models. 
For the first category, using a multiscale expansion, \cite{pro07} found a
quadratic dependence of growth rate on shear rate.
This result is contradicted by \cite{KR08} who found that the mean field
averaged over $\alpha$ does not show a dynamo and that the
fluctuating $\alpha$ effect adds to the diagonal components
of the turbulent magnetic diffusivity tensor.
\cite{HMS11} have recently proposed a
simple analytically tractable model,
in which they consider the first and second moments (calculated over
the distribution of $\alpha$) of the horizontally averaged mean field.  
The first moment shows the absence of dynamo effect, but 
the second moment shows exponential growth with
the same scaling behaviour observed by \cite{YHSKRICM07}.
Motivated by their study, we solve here a similar model using a 
technique known as {\it Gaussian integration by parts} to obtain the 
same results.
We further show that, even in the absence of shear,
it may be possible for incoherent $\alpha^2$ dynamos to operate.
For the model of \cite{VB97} we can also calculate the higher (third 
and fourth) order moments of the magnetic field to demonstrate that the
probability distribution function (PDF) of the mean magnetic field generated by
an incoherent $\alpha$--shear dynamo is non-Gaussian.  
\section[]{Mean-field model}
Our mean-field model is designed to describe the simulations of
\cite{YHSKRICM07} and BRRK.
In particular, there is a large-scale velocity given by $\Ubar= Sx\ey$.
The turbulence is generated by an isotropic external random force with zero
net helicity. 
The mean fields are defined by averaging over two coordinate directions
$x$ and $y$.
By assuming shearing-periodic boundary conditions, as done in the simulations,
this averaging obeys the Reynolds rules.
By virtue of the divergence-less property of the mean magnetic
field, v.i.z., $\partial_j \bar{B}_j = 0$, and in the absence
of an imposed mean field, $\bar{B}_z$ can be set to zero.
The resultant mean-field equations then have the following form (BRRK)
\begin{equation}
\partial_t \Bx=-\ayx\delz \Bx - \ayy\delz \By
-\etayx\del2z \By + \etayy\del2z \Bx,
\label{eq:bx}
\end{equation}
\begin{equation}
\partial_t \By=S\Bx+\axx\delz \Bx + \axy\delz \By
-\etaxy\del2z \Bx + \etaxx\del2z \By.\;
\label{eq:by}
\end{equation}
Here, $\Bx$ and $\By$ are the $x$ and $y$ components of the mean (averaged
over $x$ and $y$ coordinate directions) magnetic field,
$\eta_{ij}$ are the 
four relevant components of the turbulent magnetic diffusivity tensor and 
$\alpha_{ij}$ are the four relevant components of the $\alpha$ tensor.
The shear--current effect works via a non-zero $\etayx$,
provided its sign is the same as that of $S$.
In this paper we do not consider the possibility of the shear--current effect;
hence we set $\etayx=0$. 
Here we have ignored the molecular diffusivity,
as we are interested in the limit of very high magnetic Reynolds numbers.

The fluctuating $\alpha$ effect is modelled by choosing each 
component of the $\alpha$ tensor to be an independent Gaussian random
number with zero mean and the following covariance
(no summation over repeated indices is assumed):
\begin{equation}
\langle \alpha_{ij}(t)\alpha_{kl}(\tprime) \rangle  =
      D_{ij}\delta_{ik}\delta_{jl}\delta(t-\tprime).
\label{eq:acovar}
\end{equation} 
We also assume that the $\alpha_{ij}$s  are constants in space.
Note that DNS studies of BRRK have shown that the coefficients of turbulent
diffusivity also show fluctuations in time, but we have ignored that 
in this paper.  
To make our notation fully transparent, let us clearly distinguish between
two different kinds of averaging we need to perform. 
The mean fields themselves are constructed by Reynolds averaging, which in
our case is horizontal averaging, and is denoted by an overbar.
As we are dealing with mean field models, the quantities appearing in our
equations have already been Reynolds averaged. 
But, as the $\alpha_{ij}$ in our mean field equation are stochastic, we 
study the properties of our model by writing down evolution equations
for different moments of the mean magnetic field averaged over the statistics of 
$\alpha_{ij}$. This is denoted by the symbol $\langle \cdot \rangle$. 
To give an example, the first moment (mean over
statistics of $\alpha$) of the $x$ component of the mean magnetic field  
is denoted by $\langle \Bx \rangle$ and the second moment (mean square) is
denoted by $\langle \Bx^2 \rangle$. 

In numerical studies, such averaging has to be performed by averaging
over sufficiently many simulations with independent realisations of
noise, as was done by \cite{SS09}.
In controlled experimental setup such ensemble averaging is done by
executing the experiment several times with idential initial conditions. 
In nature we expect the noise to have a finite correlation length. 
Averaging over different realisation of the noise can then be done
by averaging over spatial domains of size larger than the correlation length
of the noise. 
Note further that in the present case the components of $\bar{\bm{B}}$
undergo a random walk and are hence non-stationary. 
Therefore we cannot replace ensemble averages by time averages.

\section{Results}
\subsection{First and second moments}
It is convenient to study the possibility of a dynamo effect in Fourier space. 
Under Fourier transform,
\begin{equation}
\Bhx = \int \Bx e^{-ikz} dk, \quad
\Bhy = \int \By e^{-ikz} dk.
\end{equation}
\Eqs{eq:bx}{eq:by} transform to,
\begin{equation}
\partial_t \Bhx=-\ayx i\kz \Bhx - \ayy i\kz \Bhy
+\etayx\kz^2 \Bhy - \etayy\kz^2 \Bhx,
\label{eq:bqx}
\end{equation}
\begin{equation}
\partial_t \Bhy=S\Bhx+\axx i\kz \Bx + \axy i\kz \Bhy
+\etaxy \kz^2 \Bhx - \etaxx \kz^2 \Bhy.\;
\label{eq:bqy}
\end{equation}
\Eqs{eq:bqx}{eq:bqy} are a set of coupled stochastic differential equations (SDEs)
with multiplicative noise.

We now want to write down an evolution equation for 
$\bm{C}^1=(\langle \Bhx \rangle ,\langle \Bhy \rangle)$,
which is the mean field averaged over the statistics of $\alpha_{ij}$.
Note that $\bm{C}^1$ is the first order moment of the probability 
distribution function (PDF) of the magnetic field. 
As the $\alpha$ effect is taken to be Gaussian and white-in-time
it is possible to obtain closed equations of the form  
\begin{equation}
\partial_t\bm{C}^1={\sf\bm{N}_1}\bm{C}^1
\label{eq:dtC1}
\end{equation}
with
\begin{equation}
{\sf\bm{N}_1}=
\left[ \begin{array}{cc}
-\ksqr(\etayy+D_{yx}) & \ksqr\etayx \\
S+\ksqr\etaxy & -\ksqr(\etaxx+D_{xy}) \end{array} \right].
\label{eq:avb}
\end{equation}
If there is no shear--current effect, i.e.\ $\etayx=0$, there is no dynamo. 
The fluctuations of the $\alpha$ effect actually enhance
the diagonal components of the turbulent magnetic diffusivity.
This result agrees with \cite{HMS11}
who used a different model and found that the mean magnetic field does not
grow. \cite{VB97} also found the same for their simplified zero-dimensional
model. 
However, if one assumes that different components of the $\alpha$ tensor
are correlated, a different result can be obtained; see 
Section~\ref{alpha-mutual}.

We have used two different techniques to derive the matrix ${\sf\bm{N}_1}$ 
in \eq{eq:avb}.
In Appendix~\ref{appendix3}, following \cite{bri+fri74}, we have used a perturbation 
expansion in powers of noise strength.
This expansion works even when $\alpha$ is {\it not} white in time.
However, for white-in-time $\alpha$ it is enough to retain only the leading
order term. 
In Appendix~\ref{appendix1} we have used {\it Gaussian integration
by parts} which works because of the white-in-time nature of the $\alpha$ effect.
This method can be easily applied to  study higher order moments of the PDF 
of the magnetic field too. Hence we shall use it extensively in the rest of
this paper. 

Although the mean magnetic field does not grow, the mean-squared magnetic field
can still show growth. To study this we now write a set of equations
for the time evolution of the second moment (covariance) of the mean magnetic 
field averaged over $\alpha_{ij}$. 
We emphasise that we are not considering here the covariance of the
actual magnetic field that would include also the small-scale magnetic
fluctuations, which are relevant to the small-scale dynamo \citep{Kaz68}.
The covariance of the actual magnetic field was later also considered
by \cite{Hoy87} in connection with $\alpha$--$\Omega$ dynamos.
Following \cite{HMS11}, we define a covariance vector, 
\begin{equation}
\bm{C}^2\equiv
(\Bhxx, \Bhyy , \Bhxy + \Bhyx ).
\end{equation}
The evolution equation for the covariance vector is given by 
\begin{equation}
\partial_t \bm{C}^2={\sf\bm{N}_2}\bm{C}^2,
\label{eq:eig_covar}
\end{equation}
where ${\sf\bm{N}_2}$ is given by
\begin{eqnarray}
\left[ \begin{array}{ccc}
-2\ksqr\etayy & 2\ksqr D_{yy} &  \ksqr\etayx \\
2\ksqr D_{xx} & -2\ksqr\etaxx & S+\ksqr\etaxy \\ 
2(S+\ksqr\etaxy) & 2\ksqr\etayx & -\ksqr(D_{yx}+D_{xy}+2\etaxx) \end{array} 
 \right].
\nonumber
\end{eqnarray}
The only non-trivial terms in the derivation of the above equation are the
terms which are products of components of $\alpha$ effect and two components
of the magnetic field. We evaluate them by using the same technique used to 
obtain \eqs{eq:dtC1}{eq:avb};  see Appendix~\ref{appendix1} for details. 

The characteristic equation of the matrix ${\sf\bm{N}}_2$ is a third
order equation, the solutions of which gives the three solutions
for the growth rate $2\gamma$.
For sim\-pli\-ci\-ty let us also choose $\etaxx=\etayy\equiv\etat$.
In other words, we take the turbulent magnetic diffusivity tensor
to be diagonal and isotropic.
We further note that  $D_{yx}$ and $D_{xy}$ contribute only in enhancing the
turbulent magnetic diffusivity of $C_3$, so we can safely ignore them compared to
$\etaxx$. 
With these simplifying assumptions the equation for the growth rate
reduces to
\begin{equation}
\xi^3 - 4\ksqr D_{yy}\left[S^2 + \ksqr D_{xx}\xi\right] = 0,
\label{eq:dispersion}
\end{equation}
where $\xi = 2(\ksqr\etad + \gamma)$. For large enough $S$
we can always ignore the second term inside the parenthesis of
\eq{eq:dispersion}. This gives the three roots of $\gamma$ as
\begin{equation}
\gamma = -\ksqr\etad + \left(\frac{1}{2}\ksqr D_{yy} S^2\right)^{1/3}
  (1,\omega,\omega^2),
\end{equation} 
where $(1,\omega,\omega^2)$ are the three cube roots of unity, of which
$\omega$ and $\omega^2$ have negative real parts. 
The same dispersion relation is obtained by \cite{HMS11}.
The wavenumber of the fastest growing mode, $\kpeak$,
is given by
\begin{equation}
\kpeak= |S|^{1/2}\left(\frac{D_{yy}}{54 \etad^3}\right)^{1/4}.
\end{equation}
The growth rate of the fastest growing mode is given by
\begin{equation}
\gamma = \frac{2^{1/3}}{6}\left(1-\frac{2^{1/6}}{\sqrt{3}} \right)
\left(\frac{D_{yy}}{\etad}\right)^{1/2} |S|.
\end{equation}
This is the same scaling numerically obtained by \cite{YHSKRICM07}.

\subsection{Incoherent $\alpha^2$ dynamo}
\label{AlpSquare}
Let us now consider a different case where shear is zero.
In that case, \eq{eq:dispersion} becomes a quadratic equation in 
$\xi$, 
\begin{equation}
\xi^2 - 4k^4 D_{xx} D_{yy} = 0,
\label{eq:quadratic}
\end{equation}
with solutions,
\begin{equation}
\gamma = \ksqr \left(-\etad \pm \sqrt{D_{xx}D_{yy}}\right).
\label{eq:da2}
\end{equation}
Hence, it may be possible for fluctuations of $\alpha$ to drive a 
large-scale dynamo (in the mean-square sense) even in the
absence of velocity shear. 

To summarise there are two possible dynamo mechanisms in our dynamo
model. In both of them the magnetic field grows in the mean-square
sense. The first one is an incoherent $\alpha$--shear dynamo. 
For large enough shear this is the fastest growing mode.
However, this dynamo has no oscillating modes because the modes
for which $\gamma$ have a non-zero imaginary part have negative real part.
An incoherent $\alpha^2$ dynamo mechanism also exists in this model.
The condition for excitation of a fluctuating $\alpha$--shear dynamo is
\begin{equation}
\frac{(\ksqr D_{yy} S^2/2)^{1/3}}{\ksqr\etat} >1 ,
\end{equation}
and the condition for excitation of a fluctuating $\alpha^2$ dynamo is
\begin{equation}
\frac{\sqrt{D_{xx}D_{yy}}}{\etat} >1 .
\end{equation}
The condition that a fluctuating
$\alpha^2$ dynamo is preferred compared to an $\alpha$--shear one is
\begin{equation}
\frac{4k^8D^3_{xx}D_{yy}}{S^4} > 1,
\end{equation}
or $\sqrt{2}\ksqr D_{xx}/|S|> 1$ for $D_{xx} = D_{yy}$. 

\begin{figure}
\includegraphics[width=0.9\columnwidth]{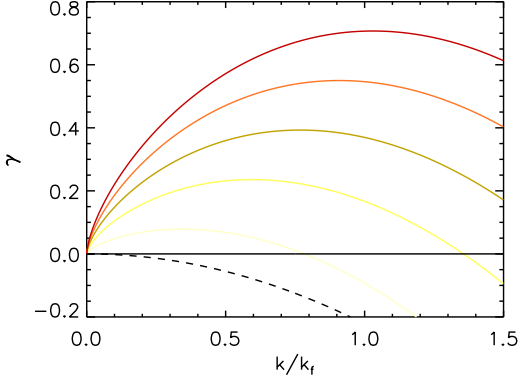}
\caption{Sketch of the dispersion relation, \eq{eq:dispersion} for different
values of velocity shear $S$ (continuous lines with difference
colours/grey shades
from bottom to top $S=0.5,1.,1.5,2.,2.5$), 
and \eq{eq:da2} (broken line) at the very bottom. Velocity shear and $\gamma$
have dimensions of inverse time and are measured in the units of  $\urms\kf$.}
\label{fig:dispersion}
\end{figure}

To compare with DNS we need to use some estimates of $\etad$, $D_{xx}$ and $D_{yy}$.
We use  $\etad = \urms/3\kf$, as obtained by \cite{Sur08} without shear.
A slightly larger value was found by BRRK in the presence of shear.
We further use $D_{xx} = D_{yy} = \urms^2/9$.
For this choice of parameters the incoherent $\alpha^2$ dynamo does not grow. 
Here, $\urms$ is the mean-squared velocity and $\kf$ corresponds to the
characteristic Fourier mode of the forcing if the turbulence has been
maintained by an external force, as done by \cite{YHSKRICM07} or BRRK.
For turbulence maintained by convection, $\kf$ should be replaced by
the Fourier mode corresponding to the integral scale of the turbulence.  
Typically, mean-field theory applies for modes with $k < \kf$.
Lengths are measured in units of $1/\kf$ and velocity is measured in the
unit of $\urms$. This makes $1/\urms\kf$ the unit of time. 
The two dispersion relations are plotted in Fig.~\ref{fig:dispersion} for
different values of $S$.
\subsection{Effects of mutual correlations between components of the $\alpha$ tensor}
\label{alpha-mutual}
So far we have assumed that only the self-correlations of the components of
the $\alpha$ tensor are non-zero and the mutual correlations zero. 
Let us now generalise \eq{eq:acovar} to 
\begin{equation}
\langle \alpha_{ij}(t)\alpha_{kl}(\tprime) \rangle  =
      \D{ij}{kl}\delta(t-\tprime).
\label{eq:acovarn}
\end{equation}
Obviously, $\D{ij}{kl} = \D{kl}{ij}$ and $\D{ij}{ij} = D_{ij}$.  
It is again possible to write a closed equation for the first moment of the
magnetic field in the form $\partial_t \bm{C}^1={\sf\bm{N}_1}\bm{C}^1$ with 
\begin{equation}
{\sf\bm{N}_1}= \left[ \begin{array}{cc}
-\ksqr(\etat+\eyy) & \ksqr\eyx \\
S+\ksqr\exy & -\ksqr(\etat+\exx) \end{array} \right]\!,
\label{eq:avbn}
\end{equation}
where
\begin{eqnarray}
\eyy = \D{yx}{yx}-\D{yy}{xx},  && \eyx = -\D{yx}{yy}+\D{yy}{xy},\\
\exy = -\D{xy}{xx}+\D{xx}{yx}, && \exx = \D{xy}{xy}-\D{xx}{yy}. 
\end{eqnarray}
This is a generalization of \eq{eq:avb}. 
Here, for simplicity, we have assumed $\etaxx = \etayy = \etat$
and $\etayx=\etaxy=0$;
in other words the conventional shear--current effect is taken to be zero. 

Note that, unlike the self-correlation terms, i.e.\ $\D{ij}{ij}$, which must be 
positive, the mutual correlation terms (e.g., $\D{yy}{xx}$, or $\D{yx}{yy}$) can have 
either sign. 
Hence, it is a-priori not clear from \eq{eq:avbn} whether a dynamo is possible or not.
However, we note two interesting possibilities below. 
First, the mutual correlations of the fluctuating $\alpha$ now contribute
to off-diagonal components of the turbulent magnetic diffusivity tensor. 
A dynamo is possible if 
\begin{equation}
\frac{S\eyx}{\ksqr\etat} > 1.
\end{equation}
Such a dynamo might look deceptively similar to the shear--current dynamo, 
but is actually
not so because in the regular shear--current effect $\etayx$ emerges due
to the presence of shear and hence must be proportional to $S$ for small $S$. 
Hence for a regular shear--current dynamo we would have
$\gamma \sim S^2$ and $\kpeak \sim S$; see BRRK.
But here $\exy$ emerges due to fluctuations of $\alpha$ and may be independent of
$S$, at least for small $S$. 
This would imply that $\gamma \sim S$ and $\kpeak \sim \sqrt{S}$
but this time even for the first moment of the magnetic field.
The growth rate of such a dynamo may however be quite small as it is proportional to
$\eyx$ which is the {\it difference} between two terms each of which are correlations 
between different components of the $\alpha$ tensor.  
Secondly, if $\D{yy}{xx} > \D{yx}{yx}$ the fluctuating $\alpha$ effect 
gives negative contributions to even the diagonal components of turbulent
diffusivity, which is reminiscent of the result of \cite{kra76}.

\subsection{Scaling in a simpler zero-dimensional model}
The essential physics of \eqs{eq:bx}{eq:by} can be captured by an even 
simpler mean-field model in a one-mode truncation, but with fluctuating $\alpha$
effect, introduced by \cite{VB97}. Their model, rewritten in our notation
and setting all $k$ factors to unity is
\begin{eqnarray}
\partial_t \Bx &=& \alpha \By - \etat \Bx, 
\label{eq:VB97a} \\
\partial_t \By &=& -S \Bx - \etat \By.
\label{eq:VB97}
\end{eqnarray}
This model can be analysed in exactly similar ways.
By construction, this model does not have a fluctuating $\alpha^2$ 
effect, and $\alpha$ is a Gaussian random variable with zero mean
and covariance
\begin{equation}
\langle \alpha(t)\alpha(\tprime)\rangle = D \delta(t-\tprime).
\label{eq:aVB}
\end{equation}
For this model we adopt a more general framework and define the growth-rate of
the $p$-th order moment of the magnetic field (the first order
is the mean and the second order is the covariance ) to be
$p\gamma_p$. 
Explicit calculations, shown in Appendix~\ref{appendix2}, give
\begin{equation}
\gamma_1 = -\etat,\quad
\gamma_2 = -\etat+\left(\frac{4DS^2}{8}\right)^{1/3} \sim S^{2/3}.
\label{eq:gamma12}
\end{equation}
We show in Appendix~\ref{appendix2} that this two-third scaling with shear
rate in this zero-dimensional model is equivalent to $\gamma \sim S$ scaling 
for \eqs{eq:bx}{eq:by}. 
\subsection{Possibility of intermittency}
\begin{figure}
\includegraphics[width=0.9\columnwidth]{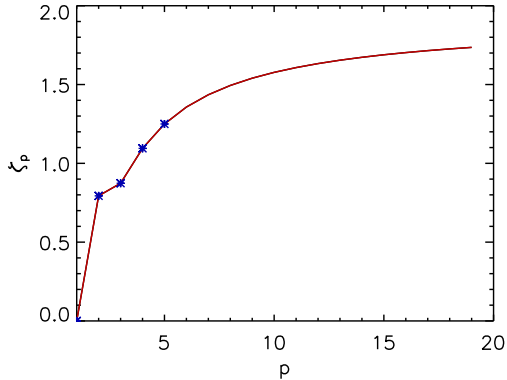}
\caption{$\zeta_p$ versus $p$ as obtained from \eq{eq:zetap}. 
If the magnetic field had obeyed Gaussian statistics, $\zeta_p$ versus
$p$ would have been constant.  }
\label{fig:zetap}
\end{figure}

We note that \eqs{eq:bqx}{eq:bqy} can be considered as coupled 
stochastic differential equations of the Langevin type but with multiplicative noise. 
We have taken the probability distribution function (PDF) of the noise to be Gaussian.
But, by virtue of multiplicative noise, the PDF of the magnetic field may
be non-Gaussian. 
We have already calculated the first and second moments
of this PDF. To probe non--Gaussianity we need to calculate the higher
order moments. For \eqs{eq:bqx}{eq:bqy} this is a formidable problem. 
But it is far simpler for the model of \cite{VB97}.
\cite{sok97} has already argued that the statistics of the magnetic field
in the model of \cite{VB97} is intermittent; see also \cite{SS09}.
In Appendix~\ref{appendix2} we show that the growth rate for the third and
fourth order moments are given by 
\begin{equation}
\gamma_3 = -\etat+\left(\frac{18DS^2}{27}\right)^{1/3}\!\!,\;\;
\gamma_4 = -\etat+\left(\frac{84DS^2}{64}\right)^{1/3}\!\!.\;\;
\label{eq:gamma34}
\end{equation}
Clearly, $\gamma_p$ has the same scaling dependence on $D$
and $S$, independent of $p$, but
nevertheless they are different, i.e., the PDF is non-Gaussian. 
This non-Gaussianity is best described by plotting
\begin{equation}
\zeta_p = \frac{\gamma_p + \etat}{(DS^2)^{1/3}}
\label{eq:zetap}
\end{equation}
versus $p$ in Fig.~\ref{fig:zetap}. 

Let us now conjecture that as $p \to \infty$, $\zeta_p$ remains finite.
Remembering that $\zeta_1 = 0$, the general form would then be
\begin{equation}
\zeta_p = \left(\frac{(p-1)(a_0+a_1p+a_2p^2)}{p^3}\right)^{1/3}.
\end{equation}
Substituting the form back in \eqs{eq:gamma12}{eq:gamma34}
we find $a_0=36$, $a_1=-30$, and  $a_2=7$.
This formula is also plotted in Fig.~\ref{fig:zetap}.
\section{Conclusions}
In this paper we have analytically solved a mean-field dynamo model with 
fluctuating $\alpha$ effect to find self-excited solutions.
We have studied the growth rate of different moments (calculated over
the statistics of $\alpha$) of the magnetic field. 
There are three crucial aspects in which our results, the DNS
of \cite{YHSKRICM07}, and the analytical results by \cite{HMS11}
agree: (a) There is no dynamo for the first moment of the magnetic field,
(b) the second moment (mean-square) of the magnetic field
shows dynamo action, and (c) the fastest growing mode has a growth
rate $\gamma \sim S$ at Fourier mode $\kpeak \sim \sqrt{S}$.
We have further shown that these aspects of our results can even
be reproduced by a simpler zero-dimensional mean field model due to
\cite{VB97}. 
For this simpler model we have also calculated the growth rate for
third and fourth order moments and we have explicitly demonstrated the
non-Gaussian nature of the PDF of the magnetic field. 
Given that the incoherent $\alpha$--shear dynamo (often with 
an additional coherent part) is the most common dynamo mechanism 
our results provide a qualitative reasoning of why large-scale
magnetic fields in the universe may be intermittent.
However note also that we have merely shown that the growth rates of the
different moments of the magnetic field are different. 
The eventual nature of the PDF of the magnetic field will also be
influenced by the saturation of this dynamo which is outside
the realm of this paper. 
We have also shown that it is possible to find growth of the first moment
(mean) of the magnetic field if mutual correlations between different components
of the $\alpha$ tensor are assumed to have a certain form (Section~\ref{alpha-mutual}).
It will thus be important to check such assumptions from future DNS.

As our paper has been inspired by \cite{HMS11} it is appropriate that
we compare and contrast our model and techniques with theirs. 
Their model consists of the equations of magnetohydrodynamics (MHD)
with an external Gaussian, white-in-time force (in the evolution equation for velocity)
with the additional assumption that the non-linear term in the velocity
equation is omitted. 
The model thus applies in the limit of Reynolds number $\Rey \ll 1$. 
They perform averaging over $xy$-coordinates to obtain mean field equations
with an $\alpha$ effect which depends on the helicity averaged over coordinate
directions. 
Our mean field model is derived by first averaging over coordinate directions
(standard Reynolds averaging) with the additional assumptions on the statistics 
(Gaussian, white-in-time) of $\alpha$.
Our results are thus not limited by the smallness of the Reynolds number,
although all the usual limitations of mean-field theory apply.
The assumption of the Gaussian nature of $\alpha$ is well supported by 
numerical evidence; see Fig.~10 of BRRK.
\cite{HMS11} have further used a quasi-two-dimensional velocity field,
but this we feel is not an important limitation. 
They average the first and second moment of the magnetic field over the 
realisations of force by using cumulant expansion in powers of the Kubo number.
As they truncate the expansion at the lowest order in Kubo number it applies
to the case of small Kubo number.  
The most restrictive assumption in our model is the assumption of the
white-in-time nature of the $\alpha$ effect. 
This assumption however allows us to obtain closed equations for 
all the moments of the magnetic field.
The results of \cite{HMS11} is not limited by this assumption.
It is interesting to note that, even under the assumption
of the white-in-time nature of the $\alpha$ effect, we obtain the same
scaling behaviour as \cite{HMS11} and the DNS studies of \cite{YHSKRICM07}.

Here, let us also mention that  
\cite{kol+leb+siz10} have recently applied similar techniques to study
{\it small-scale} kinematic dynamos in a smooth delta-correlated velocity field 
in the presence of shear to find
\begin{equation}
\gamma_n = \frac{3}{2^{5/3}}n^{4/3}D^{1/3}S^{2/3} \sim \lambda n^{4/3},
\end{equation}
where $\lambda$ is the expression for the largest Lyapunov exponent 
describing the divergence of two initially close fluid particles, 
This Lyapunov exponent was earlier obtained for such flows by \cite{tur07}. 
Interestingly, this is exactly the same scaling with shear as in \cite{VB97}.
In the absence of shear the small-scale dynamo can still 
operate \citep{che+kol+ver97} with $\gamma_n \sim n^2$. 

\cite{pro07} have also considered a model similar to ours, although
somewhat simpler and more relevant to the solar dynamo,
using multiscale expansions.
After averaging over the fluctuating $\alpha$ effect, he still finds
an effective $\alpha$ effect from which he obtains a dynamo
which grows in the mean (as opposed to mean--square in our case).
His results give the scaling, $\kpeak \sim S$ and $\gamma \sim S^2$ 
which disagree with the DNS results of \cite{YHSKRICM07}.
\section*{Acknowledgements}
We thank Tobi Heinemann, Matthias Rheinhardt, and Alex Schekochihin
for useful discussions on the incoherent $\alpha$--shear dynamo.
Financial support from the European Research Council under the AstroDyn
Research Project 227952 is gratefully acknowledged. 


\appendix

\section[]{Derivation of equations (7) and (8)}
\label{appendix3}
It is possible to derive \eq{eq:avb} via a method described 
by \cite{bri+fri74}. 
This method is superior to the one described in Appendix \ref{appendix1}
in the sense that this can be applied even when $\alpha$ is not necessarily 
white in time, but has (small) non-zero correlation time. 
On the other hand it is more cumbersome to apply this method to calculate
the higher order moments of the PDF of the magnetic field. 
For the sake of completeness we reproduce below the calculations of \cite{bri+fri74}
as applied to our problem. 

Let us write symbolically the evolution equations for the mean-field in Fourier space
in the following way,
\begin{equation}
\partial_t \cB = \left[ \MMd + \MM \right] \cB .
\label{eq:mat}
\end{equation}
Here, $\cB = (\Bhx, \Bhy)$ is a column vector,
$\MMd$ is the deterministic part of the evolution equations
(i.e., the part that depends on $\eta_{ij}$),
\begin{equation}
\MMd=
 \left[ \begin{array}{cc}
-\ksqr\eta_{yy} & 0 \\
S & -\ksqr\eta_{yy} \end{array} 
 \right],
\end{equation}
which is also independent of time, 
and $\MM$ is the random part (i.e., the part that 
depends on $\alpha_{ij}$), with
\begin{equation}
\MM=
 i\kz\left[ \begin{array}{cc}
-\ayx  & -\ayy \\
\axx & \axy \end{array} 
 \right] \equiv i\kz \AA.
\label{eq:MM}
\end{equation}
To begin, we do not assume that $\alpha$ is white-in-time but that it has finite
correlation time $\tcor$. Later we shall take the limit of $\tcor \to 0$ 
in a specific way to reach the white-noise-limit. 
This is equivalent to the regularization in Appendix \ref{appendix1}

In this section, for simplicity, let us choose our units such that at $t=0$, 
$\cB = (1,1)$. 
In that case the solution to \eq{eq:mat} can be easily recast in the
integral form
\begin{equation}
\cB(t) = e^{\MMd t} + 
      \int_0^{t} d\tprime e^{\MMd(t-\tprime)}\MM(\tprime)\cB(\tprime).
\label{eq:integral}
\end{equation}
Note that we have $\langle \MM \rangle = 0 $ because we have assumed all the components 
of $\alpha$ to have zero mean.
Iterating this equation, we obtain
\begin{eqnarray}
&&\!\!\!\cB(t) = e^{\MMd t} + 
       \int_0^{t}d\tprime e^{\MMd(t-\tprime)}\MM(\tprime)e^{\MMd \tprime} 
         \label{eq:pert}   \\
       &&\!\!\!+ \int_0^{t}d\tprime\int_0^{\tprime} d\tpp 
      e^{\MMd(t-\tprime)}\MM(\tprime) e^{\MMd(\tprime-\tpp)}\MM(\tpp)\cB(\tpp)
\nonumber
\end{eqnarray}
$+\mbox{higher order terms}$.
Let us also assume that the strength of the fluctuations of $\alpha$ are finite
and bounded by $\sigma$. 
\Eq{eq:pert} is then an expansion in powers of $\sigma t$. 
To obtain closed equations for the first moment of the magnetic field,
we average \eq{eq:pert} over the statistics of $\alpha$ and then take
the derivative with respect to $t$. 
Remembering our earlier notation ${\bm C}^1 \equiv \langle \cB \rangle$,
we obtain
\begin{equation}
\partial_t {\bm C}^1(t) = \MMd {\bm C}^1(t) +
     \int_0^{t}d\tprime 
     \langle \MM(t) e^{\MMd(t-\tprime)}\MM(\tprime)\cB(\tprime) \rangle.
\label{eq:master}
\end{equation}
This obviously is not yet a closed equation. 
To obtain a closure, note that for 
\begin{equation}
\sigma(\tprime -s) \ll 1,
\label{eq:kubo1}
\end{equation} 
from \eq{eq:integral}, we have
\begin{equation}
\cB(\tprime) \approx e^{\MMd(\tprime -s)}\cB(s) + O(\sigma(\tprime -s)).
\label{eq:cBt}
\end{equation}
Substituting \eq{eq:cBt} in the integrand of the double integral in 
\eq{eq:master} we obtain the factorization,
\begin{eqnarray}
\langle \MM(t) e^{\MMd(t-\tprime)}\MM(\tprime)e^{\MMd(\tprime -s)}\cB(s) \rangle
\approx \nonumber \\ 
\langle \MM(t) e^{\MMd(t-\tprime)}\MM(\tprime)\rangle 
 \langle e^{\MMd(\tprime -s)}\cB(s) \rangle,
\label{eq:clo1}
\end{eqnarray}
if we assume that 
\begin{equation}
\tprime -s \gg \tcor.
\label{eq:kubo2}
\end{equation}
\Eqs{eq:kubo1}{eq:kubo2} can both hold only
for small Kubo number, 
\begin{equation}
K = \sigma \tcor \ll 1 .
\label{eq:kubo}
\end{equation}
Substituting \eq{eq:clo1} in \eq{eq:master} and using again \eq{eq:cBt}
we obtain 
\begin{equation}
\partial_t {\bm C}^1(t) = \MMd {\bm C}^1(t) +
     \int_0^{t}d\tprime 
     \langle \MM(t) e^{\MMd(t-\tprime)}\MM(\tprime)\rangle 
       {\bm C}^1(\tprime),
\label{eq:bourret}
\end{equation}
Following \cite{bri+fri74}, we shall call this equation the Bourret equation.
For small Kubo number we can invert \eq{eq:cBt} to have
\begin{equation}
\cB(\tprime) \approx e^{-\MMd(t-\tprime)}\cB(t) .
\label{eq:invert}
\end{equation}
Averaging \eq{eq:invert} over the noise we obtain
\begin{equation}
{\bm C}^1(\tprime) \approx e^{-\MMd(t-\tprime)}{\bm C}^1(t) . 
\label{eq:Cinvert}
\end{equation}
Substitute this back into \eq{eq:bourret}, noting in addition
that for short-correlated $\alpha$ and  $t \gg \tcor$, we
can replace $\tau \equiv t - \tprime$ in \eq{eq:bourret} and
extend the integral from zero to infinity to obtain
\begin{equation}
\partial_t {\bm C}^1 = \MMd {\bm C}^1 +
     \int_0^{\infty}d\tau 
     \langle \MM(\tau) e^{\MMd\tau}\MM(0)e^{-\MMd\tau}\rangle {\bm C}^1,
\label{eq:clo2}
\end{equation}
where we have omitted the $t$ argument on all ${\bm C}^1(t)$ for brevity.
To get this result, remember that the integral above gives negligible contribution
for $\tau \gg \tcor$ and the matrices $\MMd$ and $\MM(\tau)$ do not 
necessarily commute.  
To go to the white-in-time limit we need to take the limit $\sigma \to \infty$,
$\tcor \to 0$ in such a way that the product $\sigma^2 \tcor$ remains finite.
In this limit the Kubo number goes to zero and the various approximations
made above become exact.  
In this limit the integral in \eq{eq:clo2} reduces to
\begin{eqnarray}
\partial_t {\bm C}^1(t) &=& \MMd {\bm C}^1(t) + \langle \MM \MM \rangle {\bm C}^1(t)
 \nonumber \\
&=& \left[ \MMd  -\ksqr \langle \AA \AA \rangle \right] {\bm C}^1(t).
\label{eq:clo3}
\end{eqnarray}
The correlator on the right hand side of \eq{eq:clo3} can be obtained by 
using \eq{eq:MM} together with \eq{eq:acovar}. This reproduces \eq{eq:avb}.
If instead of \eq{eq:acovar}, \eq{eq:acovarn} is used, \eq{eq:avbn} can be obtained.

Finally, note that for an $\alpha$ effect which has finite-time-correlations (instead of white-in-time)
higher order terms in the expansion in \eq{eq:pert} are needed. 
In such cases it may not even be possible to obtain closed equations like \eq{eq:dtC1}.

\section[]{Averaging over Gaussian noise}
\label{appendix1}
We explain here another technique used to derive \eqs{eq:dtC1}{eq:avb}.
Let us begin by considering Gaussian vector-valued noise $\nu_j(t)$ 
(not necessarily white-in-time)
and an arbitrary functional of that, $F(\bm{\nu})$. Then,
\begin{equation}
\langle F(\bm{\nu})\nu_j(t) \rangle = \int d\tprime \langle \nu_j(t)\nu_k(\tprime)\rangle
                               \left \langle \fder{F}{\nu_k(\tprime)}
                                 \right \rangle,
\label{eq:gaussian}
\end{equation}
where the average factorises by virtue of the Gaussian property of the noise.
Here the operator $\fdertxt{\cdot}{\nu_k}$ is the functional derivative
with respect to $\bm{\nu}$.
This useful identity often goes by the name {\it Gaussian integration by parts};
see, e.g., \cite{Zin99}, Section 4.2  for a proof; see also e.g., \cite{Fri96},
\cite{fri+wir97}, or \cite{mit+pan05}, where this method has been used to derive 
closed moment equations for the Kraichnan model of passive scalar 
advection \citep{kra68}. 

To obtain an evolution equation for 
$\bm{C}^1=(\langle \Bhx \rangle ,\langle \Bhy \rangle)$ 
we average each term of \eqs{eq:bqx}{eq:bqy} over the statistics of 
$\alpha_{ij}$. 
Terms which are product of components of $\alpha_{ij}$ and $\Bhx$ or
$\Bhy$ can be evaluated by using the identity in \eq{eq:gaussian}. 
In particular,
\begin{eqnarray}
\langle \ayx \Bhx \rangle &=&  
 \int d\tprime \langle \ayx(t)\alpha_{kl}(\tprime) \rangle
            \left \langle \fder{\Bhx(t)}{\alpha_{kl}(\tprime)} \right \rangle 
\nonumber \\
&=& D_{yx} \fder{\Bhx(t)}{\alpha_{kl}(t)}.
\label{eq:alpha_b_avg}
\end{eqnarray}
Here we have considered the magnetic field to be a functional of $\alpha_{ij}$.
Substituting the covariance of $\alpha_{ij}$ from \eq{eq:acovar}, 
integrating the $\delta$ function over time and contracting over the Kronecker deltas,
we obtain the last equality in \eq{eq:alpha_b_avg}. 

The functional derivatives of the components of the magnetic field
with respect to $\alpha_{ij}$ can be obtained by first formally integrating 
\eqs{eq:bqx}{eq:bqy} to obtain $\Bhx(t)$ and $\Bhy(t)$, respectively,
and then calculating their functional derivatives with respect to $\alpha_{ij}$.
We actually need the functional derivative $\fdertxt{\Bhx(t)}{\alpha_{kl}(\tprime)}$
and then take the limit $t\to\tprime$.
This is a non-trivial step due to the singular nature of the correlation function
of $\langle\alpha_{kl}(t)\alpha_{kl}(\tprime)\rangle$ as $t\to\tprime$. 
To get around the difficulty it is possible to replace the Dirac delta function
in \eq{eq:acovar} with a regularised even function and then take limits. 
We refer the reader to \cite{Zin99}, Section 4.2, for a detailed discussion.
This regularization is equivalent to using the Stratanovich prescription 
for the set of coupled SDEs \eqs{eq:bqx}{eq:bqy}; see, e.g., \cite{Gar94}.
For reference, all the non-zero functional derivatives needed are given below:
\begin{eqnarray}
\fder{\Bhx(t)}{\alpha_{yx}(t)} = -i\kz \Bhx , &&  
\fder{\Bhx(t)}{\alpha_{yy}(t)} = -i\kz \Bhy , \nonumber \\
\fder{\Bhy(t)}{\alpha_{xx}(t)} =  i\kz \Bhx , &&
\fder{\Bhy(t)}{\alpha_{xy}(t)} =  i\kz \Bhy .
\label{eq:fder}
\end{eqnarray}
In particular, since there is no $\alpha_{yy}$ term
in \eq{eq:bqy}, we have $\fdertxt{\Bhy(t)}{\alpha_{yy}(t)} = 0$, so
\begin{equation} 
\left\langle \ayy  \Bhy \right\rangle
= D_{yy} \left\langle \fder{\Bhy}{\ayy} \right\rangle
= 0.
\nonumber
\end{equation}
Putting everything together we can now average \eq{eq:bqx} over the statistics
of $\alpha$ to obtain,
\begin{eqnarray}
\partial_t \langle \Bhx \rangle &=& 
-\langle i\kz \ayx  \Bhx \rangle  - \langle i\kz \ayy  \Bhy \rangle \nonumber \\
&&+\etayx\kz^2 \langle \Bhy \rangle - \etayy\kz^2 \langle \Bhx\rangle \nonumber \\
&=& -i\kz(-i\kz)D_{yx} \langle \Bhx \rangle  \nonumber \\
&&+\etayx\kz^2 \langle \Bhy \rangle - \etayy\kz^2 \langle \Bhx\rangle \nonumber \\
&=& -(\etayy+D_{yx})\ksqr \langle \Bhx \rangle  + \etayx\kz^2 \langle \Bhy \rangle.
\end{eqnarray}
This gives us the first row of the matrix ${\sf\bm{N}_1}$ in \eq{eq:avb}.
Applying the same technique to \eq{eq:bqy}, the second row can be obtained.

The same technique can be applied to obtain the matrix ${\sf\bm{N}_2}$.
Here we end up with evaluating terms of the general form
\begin{equation}
\langle \alpha_{ij} \hat{B}_{k}\hat{B}^{\ast}_{l}\rangle 
 = D_{ij}\left\langle \fder{\hat{B}_{k}}{\alpha_{ij}}\hat{B}^{\ast}_{l} + 
          \hat{B}_{k}\fder{\hat{B}^{\ast}_{l}}{\alpha_{ij}}\right\rangle.
\end{equation}
The only non-zero functional derivatives are given in \eq{eq:fder}.
We show the calculation explicitly for the two following examples:
\begin{eqnarray}
\langle \alpha_{yx}\bhxx \rangle &\!\!=\!\!& D_{yx}\left\langle\fder{\Bhx}{\alpha_{yx}}\Bhxcc
                                      +\fder{\Bhxcc}{\alpha_{yx}}\Bhx \right\rangle
\nonumber \\
   &\!\!=\!\!& D_{yx}\left[-ik \Bhxx + ik \Bhxx \right] = 0,
\end{eqnarray}
and 
\begin{eqnarray}
\langle \alpha_{yy}\bhyx \rangle &\!\!=\!\!& D_{yy}\left\langle\fder{\Bhy}{\alpha_{yy}}\Bhxcc
                                      +\fder{\Bhxcc}{\alpha_{yy}}\Bhy \right\rangle
\nonumber \\
   &\!\!=\!\!& ikD_{yy}\Bhyy.
\end{eqnarray}

Note further that, instead of using the functional calculus above, the same
evolution equations for ${\bm C}^1$ can be obtained by using the 
technique due to  \cite{bri+fri74}.
This is demonstrated in Appendix \ref{appendix3}.

\section[]{A zero-dimensional mean-field model with fluctuating $\alpha$}
\label{appendix2}
A simpler mean-field model in a one-mode truncation, but with fluctuating $\alpha$
effect, was introduced by \cite{VB97}; see \eqs{eq:VB97a}{eq:VB97}.
For this model we define the following moments of successive orders,
\begin{eqnarray}
\bm{C}^1 &\equiv& (\langle \Bx \rangle, \langle \By \rangle), \\
\bm{C}^2 &\equiv& (\langle \Bx^2 \rangle, \langle \By^2 \rangle, \langle \Bx\By \rangle ), \\
\bm{C}^3 &\equiv& (\langle \Bx^3 \rangle, \langle \Bx^2\By \rangle, 
    \langle \Bx\By^2 \rangle, \langle \By^3 \rangle ), \\
\bm{C}^4 &\equiv& (\langle \Bx^4 \rangle, \langle \Bx^3\By \rangle, 
    \langle \Bx^2\By^2 \rangle,  \langle \Bx\By^3 \rangle, 
     \langle \By^4 \rangle ).
\end{eqnarray}
Each of these moments satisfies a closed equation of the form
\begin{equation}
\partial_t \bm{C}^p = {\sf\bm{N}_p}\bm{C}^p.
\end{equation}
The matrices ${\sf\bm{N}_p}$ can be found by applying the technique 
described in Appendix~\ref{appendix1} and by using the 
covariance of $\alpha$ given in \eq{eq:aVB}.  
We give below the first four matrices:
\begin{equation}
{\sf\bm{N}_1}=
 \left[ \begin{array}{cc}
-\etad  & 0 \\
-S & -\etat \end{array} \right],
\end{equation}
\begin{equation}
{\sf\bm{N}_2}=
 \left[ \begin{array}{ccc}
-2\etat & 2D &  0 \\
0 & -2\etat & -2S \\ 
-S & 0 & -2\etat \end{array} 
 \right],
\label{eq:cVB}
\end{equation}
\begin{equation}
{\sf\bm{N}_3}=
 \left[ \begin{array}{cccc}
-3\etat & 0 & 6D &  0 \\
-S & -3\etat & 0 & D \\ 
0 & -2S & -3\etat & 0 \\
0 & 0 & -3S & -3\etat  \end{array} 
 \right],
\label{eq:VB3}
\end{equation}
\begin{equation}
{\sf\bm{N}_4}=
 \left[ \begin{array}{ccccc}
-4\etat & 0 & 12D &  0 & 0\\
-S & -4\etat & 0 & 6D & 0\\ 
0 & -2S & -4\etat & 0 & 2D\\
0 & 0 & -3S & -4\etat & 0 \\
0 & 0 & 0   & -4S & -4\etat \end{array} 
 \right].
\label{eq:VB4}
\end{equation}
The growth rate at order $p$ is defined to be 
$\bm{C}^p \sim \exp(p\gamma_p t)$.
This gives $\gamma_1 = -\etat$, i.e., there is no dynamo. 
But this also gives dynamo modes with positive eigenvalues given by
\begin{equation}
\gamma_p \sim S^{2/3}D_{xx}^{1/3},\quad p = 2,3, \ldots
\label{eq:VBscaling}
\end{equation}
The same result was obtained by \cite{VB97} for $\gamma_2$ by using a different method. 

Note the striking similarity between matrix ${\sf\bm{N}_2}$ in \eq{eq:cVB}
and matrix ${\sf\bm{N}_2}$ in \eq{eq:eig_covar}. 
A trivial way of generalising \eq{eq:cVB} to one spatial dimension is to 
replace $D$ and $\etat$ in \eq{eq:cVB} by $\ksqr D$ and $\ksqr\etat$,
respectively. 
The solution of the resultant eigenvalue problem gives the scaling,
$\gamma \sim S$ and $\kpeak \sim \sqrt{S}$. 
Thus, \eq{eq:cVB} for this zero dimensional model is equivalent to
\eq{eq:eig_covar} in the space--time model.

\end{document}